\documentclass[%
reprint,
amsmath,amssymb,
 aps,
pre,
floatfix,
]{revtex4-2}
\usepackage{float}
\usepackage{graphicx}
\usepackage{dcolumn}
\usepackage{bm}
\usepackage{hyperref}
\usepackage[mathlines]{lineno}

\begin{document}

\preprint{APS/123-QED}

\title{\textbf{Experimental evidence of an Apolar Biaxial Smectic-A Phase Comprised of  Bent-Core   Molecules}}
\author{ Susovan Bhandary}
\author{Abhijith K}
\author{Arun Roy}%
\email{aroy@rri.res.in}
\affiliation{%
Raman Research Institute,
C. V. Raman Avenue,
Sadashivanagar, Bengaluru 560080, India}%
\date{\today}

\begin{abstract}
 We report structural and physical investigations on the biaxial smectic-A phase exhibited by a compound consisting of asymmetric bent-core molecules. Upon cooling from the isotropic phase, the compound exhibits the following phase sequence: Isotropic (403.9 K) $\rightarrow$ biaxial Smectic-A (359.8 K) $\rightarrow$ Crystal. The polarized optical microscopy, X-ray diffraction, and polarization reversal current measurements clearly establish the biaxial nature of the smectic-A phase without any layer polarization. The measured layer spacing in the entire temperature range of the smectic-A phase is close to the molecular length. The schlieren texture in a homeotropically aligned sample shows both $\pm \frac{1}{2}$ and $\pm 1$ defects which indicate the biaxial nature of this smectic-A phase. The dielectric spectroscopy studies on the samples revealed Debye-type relaxation processes with the relaxation time following the Arrhenius equation with temperature.   Interestingly, the observed biaxial smectic-A phase exhibits a remarkable electro-optic response for a planar-aligned sample without any reorganization of the smectic layer structure. 

\end{abstract}

\maketitle


\section{INTRODUCTION} Liquid crystals (LCs) are intermediate phases (mesophases) that exist between the more ordered crystalline solid phase and highly disordered isotropic liquid phase composed of strongly anisotropic molecules. By definition, LC phases retain fluidity along with some characteristics of anisotropic crystals~\cite{Chandrasekhar_1992}. The majority of  known thermotropic liquid crystal compounds are rod-like organic molecules, and they exhibit two main groups of LC phases, nematics and smectics. In the Nematic phase, the constituent rod-like molecules have only long-range orientational order without any translational order between them. The direction along which the long axes of the molecules tend to align on average is called the director $\textbf{\^{n}}$. The uniaxial nematic is the most commonly observed phase and has been exploited in liquid crystal displays. Unlike the uniaxial nematic phase, a biaxial nematic phase requires a second director $\textbf{\^{m}}$, orthogonal to $\textbf{\^{n}}$, to describe the orientational ordering of the biaxial board-like molecules. This phase was first theoretically predicted by Michael J. Freiser in 1970 \cite{freiser1970BLNC}. Since then, a few reports of the observation of biaxial nematic phase in thermotropic liquid crystals have been published~\cite{Merkel2004NbPRL,Acharya2004NbPRL,Madsen2004prlNb}. However, conclusive evidence of biaxiality in these systems still remains debatable, and despite intensive investigations, the biaxial nematic phase remained elusive in thermotropic systems~\cite{Galerne2006NbPRL,jakli2013liquid,kim2013surface}.
Smectic phases are less symmetric than nematic phases as they also possess long-range positional order in at least one spatial direction in addition to long-range orientational order. In the uniaxial Smectic-A (SmA) phases, the anisotropic LC molecules arrange themselves into liquid-like layers, with the director $\textbf{\^{n}}$ parallel to the layer normal. The  SmA phase has the layer spacing  of the order of the molecular length. Though the existence of the uniaxial SmA phase is well established, the observation of the biaxial SmA phase is comparatively rare~\cite{leube1991optical,hegmann2001evidence}. De Gennes first envisaged the possible existence of the apolar biaxial smectic-A phase in the first edition of his classic book~\cite{deGennes1974}. 
The search for novel liquid crystal compounds for electro-optic applications over the past three decades has resulted in the development of bent-core banana-shaped molecules\cite{link1997spontaneous, niori1996dSmAb}. The bent-core liquid crystals are found to
exhibit a wide range of liquid crystal phases, including several novel smectic phases characterized by spontaneous breaking of chiral symmetry~\cite{reddy2006bent,Jackli2018RMP}.
The bent-core molecules exhibiting polar biaxial SmA phase with in-plane layer polarization has also been reported \cite{Eremin2001PRE,shreenivasa2004polar,reddy2004direct,semmler1998biaxial,panarin2010field,reddy2011spontaneous, Aroy2011softmatter}. But the experimental evidence of an apolar biaxial smectic A phase has remained quite limited~\cite{meyer2021freedericksz,walker2021remarkable}.
Meyer \textit{et al.} \cite{meyer2021freedericksz} reported the detection of a biaxial Smectic-A  (SmA$_b$) phase in mixtures composed of the odd spacer dimers and a small fraction of the rod-like molecules. They also found  an electro-optic response  in the planer aligned  SmA$_b$ phase. Walker et al.  found a SmA$_b$ phase in a pure compound composed of the odd spacer dimers of rod-like moieties~\cite{walker2021remarkable}.
However, the biaxial smectic-A phase was found to have an intercalated layer structure for their system.

In this work, we describe the physical properties of an enantiotropic apolar biaxial Smectic-A (SmA$_b$) phase 
of a pure compound comprised of bent-core banana-shaped molecules. 
Unlike conventional bent-core smectic phases, the observed SmA$_b$ phase exhibits no net in-plane layer polarization.  X-ray diffraction reveals a layer spacing comparable to the molecular length, indicating a non-intercalated layer structure.  The phase also exhibits a pronounced electro-optic response in planar-aligned liquid crystal cells. The measured complex dielectric function exhibits Debye type relaxation. The temperature dependence of the relaxation time follows the empirical Arrhenius equation. To the best of our knowledge, this is the first report of a pure bent-core liquid crystal exhibiting a non-intercalated apolar biaxial Smectic-A phase that displays electro-optic response.

\section{EXPERIMENTAL}
 The phase transition temperatures of the bent-core liquid crystal was studied by differential scanning calorimetry (DSC) using a Mettler Toledo DSC 3 instrument. The sample was sealed in an aluminium crucible, and an identical empty crucible was used as the reference. Thermograms were recorded at a scanning rate of 1.5 K/min during both heating and cooling runs.
Variable-temperature X-ray diffraction (XRD) studies were performed using a DY 1042-Empyrean (PANalytical) diffractometer with  CuK$_\alpha$ radiation of wavelength 1.54 $\mathrm{\AA}$ and a PIXcel 3D detector. The sample was sealed in a Lindemann capillary tube of outer diameter 1 mm. The diffracted intensity profiles were collected at various temperatures during cooling from the isotropic phase. 

Polarized optical microscopy (POM) investigations of the sample were performed using an Olympus
BX 50 microscope equipped with a digital camera (Canon EOS 80D). The temperature of the sample was monitored using a microscope hot stage (Linkam LTS420E) and a temperature controller (Linkam-T95). The commercially available LC cells (INSTEC Inc.) of gap 5 $\mu m$ and 9 $\mu m$ were used for the planar
alignment of the sample. The LC cells are made of indium tin oxide (ITO)-coated glass plates which
act as electrodes of effective area $5\times5 \ mm^2 $ for electro-optic and dielectric studies. The sample was filled into the LC cells
by capillary action in its isotropic phase using a hot plate. It was found that the sample
did not align homeotropically in the corresponding commercially available LC cells. However,  
it aligned homeotropically when filled in a custom made ITO-coated cell, which was not treated 
for any kind of alignment of the molecules. It is also observed that homeotropic alignment of the
sample can be obtained in a planar cell by cooling it from its isotropic to Smectic phase in 
the presence of a sinusoidal electric field of rms amplitude about 4 volts/$\mu$m and frequency 
20~kHz~\cite{meyer2021freedericksz}. The homeotropic texture once obtained in the Smectic phase 
was retained even after switching off the electric field. 

The electric polarization of the sample at different
temperatures was investigated using the triangular wave voltage technique
for planar aligned samples\cite{miyasato1983direct}. The
current response of the sample was measured using an oscilloscope 
(Agilent MSO6012A) by recording the voltage drop across a 
1 k$\Omega$ resistor, which was connected in series with the sample cell.

To estimate the effective birefringence of the planar-aligned sample and its temperature 
dependence, the mean transmitted intensity through the sample was measured as a function of temperature. The planar-aligned sample cell was mounted on a rotatable microscope stage between crossed polarisers and the sample cell was oriented such that the selected region of interest exhibited maximum transmittance. In this orientation, the director in the region of interest makes an angle of 45 degrees with respect to the polarizer. Images were recorded at different temperatures after inserting a long pass red filter with a cutoff wavelength of 650 nm into the optical path. The mean transmitted intensity was subsequently calculated from the POM images using Fiji and MATLAB software. The mean transmitted intensity was used to calculate the effective birefringence of the sample.

Both commercially available LC cells and custom-made non-treated LC cells were used for dielectric measurements using two types of setup. The variation of the dielectric constant as a function
of temperature was measured using a custom-built setup. During measurement,
a sinusoidal voltage of frequency 5641 Hz and rms amplitude 0.5 V was applied to the sample cell,
which was connected with a 1 k$\Omega$ resistor in series. A lock-in amplifier(Stanford Research
SR830) was employed to measure the amplitude and phase of the voltage drop across the 1 k$\Omega$
resistor.  The capacitance of the sample cell was measured by impedance analysis. The effective
dielectric constant of the sample is obtained by dividing the capacitance of the filled cell 
by that of the empty cell. The temperature of the sample was monitored using a hot stage and a
temperature controller(Instec Inc) with a temperature precision of 0.1 degrees. Due to the limited
frequency range of the custom-built setup, the frequency dependence of the dielectric constant was
investigated using a Novocontrol Alpha analyzer which has a frequency range up to 40 MHz. In
this setup, the temperature of the sample was controlled using a hot stage and temperature 
controller(Linkam T95HS). Before filling the sample, the empty capacitance of the cell was measured,
which is used to calculate the dielectric constant of the sample.
\label{Section__Expt}
\section{RESULT AND DISCUSSION}
\subsection{Phase sequence}
\begin{figure}[!t]
\centering
\includegraphics[width=0.75\linewidth]{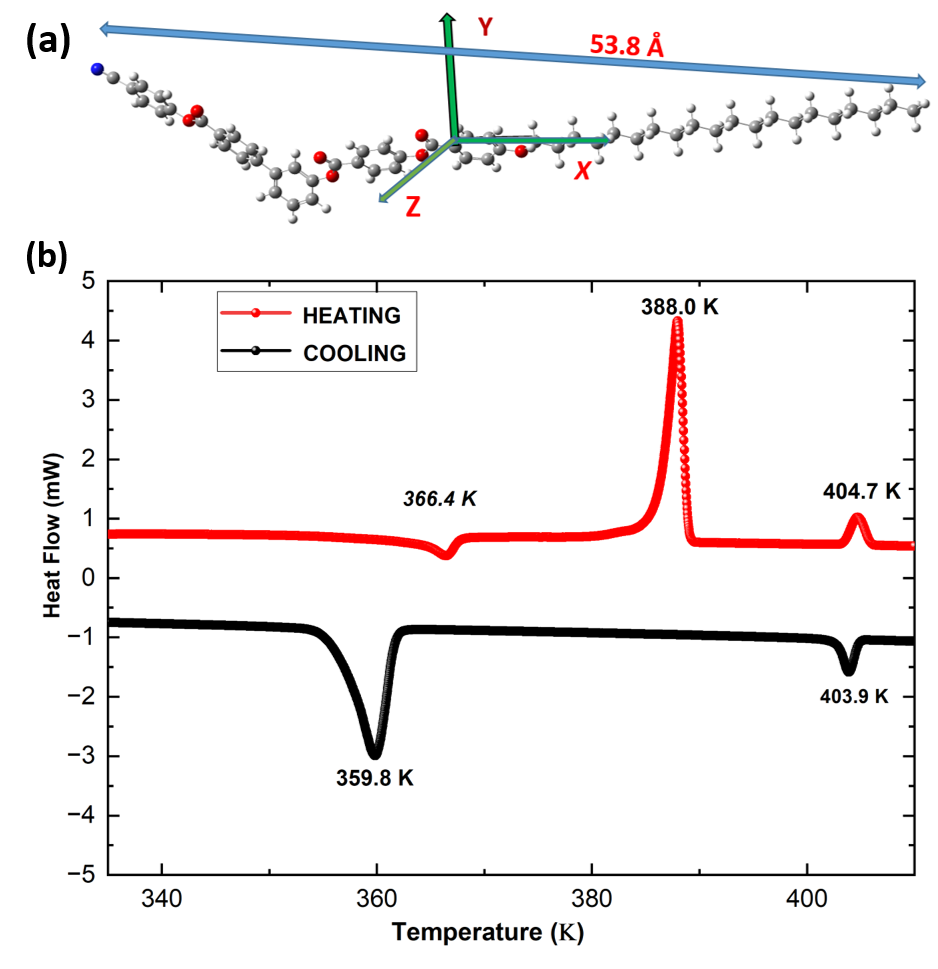}
\caption{ (a) Energy-minimised molecular structure and the molecule fixed coordinate axes used in the calculation for the compound CBBO20. (b) DSC thermogram of the sample showing the phase transition peaks on heating and cooling at the rate of 1.5 K/min}
\label{figure_DSC}
\end{figure}

The compound used in our experimental studies is \emph{4-Cyanophenyl-3'-(4-(4-20-alkoxybenzoyloxy)benzoyloxy)biphenyl-4-carboxylates} 
which will be denoted hereafter as CBBO20~\cite{Radhika2012}. 
The molecule has an asymmetric bent-core banana shape as depicted in Fig.\ref{figure_DSC}(a). 
The opening angle and the molecular length, calculated from the energy-minimized molecular 
structure, are about $ 143 ^\circ$ and 53.8 $\mathrm{\AA}$, respectively. A strongly polar carbonitrile 
group attached to one end of the molecule enhances the net dipole moment, which is directed 
almost along the molecular long axis. The components of the dipole moment calculated using 
Gaussian 16 software are  $\mu_{x} = 11.69$ Debye, $\mu_{y} = -5.36$ Debye,  
$\mu_{z} = -2.34$ Debye (see Fig.\ref{figure_DSC}(a)).
Differential scanning calorimetry (DSC) studies have been carried out on the sample  CBBO20 to detect the various phase transitions during heating and cooling cycles at a rate of 1.5 K/min. The DSC thermogram of the sample during heating and cooling run is shown  in Fig.\ref{figure_DSC}(b).  During heating, it exhibits an exothermic crystal-to-crystal transition at 366.4 K. On further heating, it melts to a mesophase at 388.0 K and undergoes a first order transition to the isotropic phase at 404.7 K. The enthalpy change for the mesophase to  isotropic phase transition is  $5.18 $ kJ/mol. During cooling from the isotropic phase, the sample goes to the mesophase at 403.9 K with an enthalpy change of $5.08$ kJ/mol and then upon further cooling, it crystallizes at 359.8 K. Therefore, the DSC observations clearly indicate the existence of only one mesophase of the sample  CBBO20. A small hysteretic shift in the isotropic to mesophase transition temperature was observed while the  mesophase to crystal transition temperature shows a large shift.
\label{subsection_DSC}
\subsection{X-ray diffraction Studies}
\begin{figure}[!t]
\centering
\includegraphics[width=0.75\linewidth]{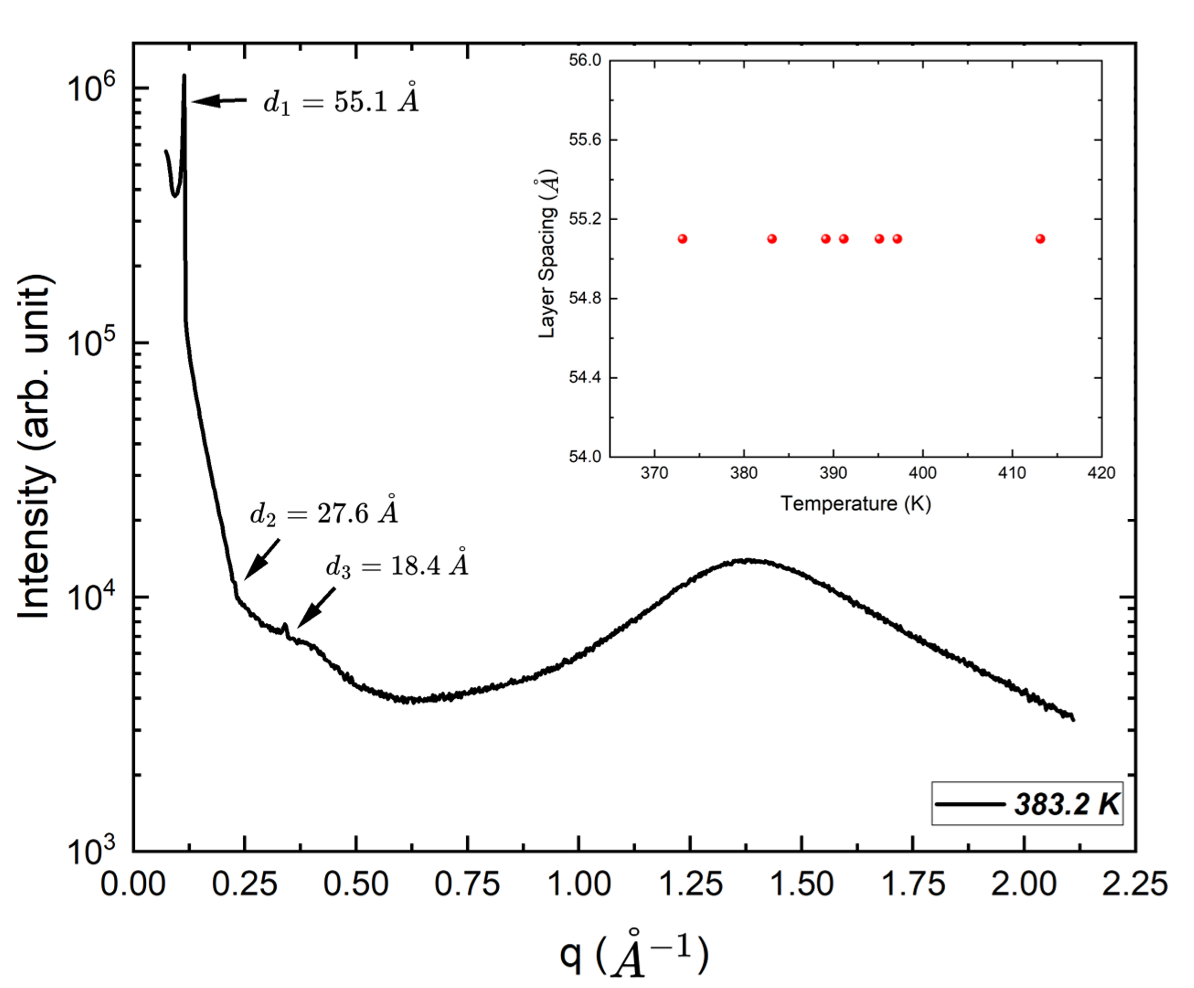 }
\newline 
\caption{ X-ray diffraction intensity profile as a function of wave-vector $q$ in the SmA$_b$ phase at 383.2 K. The inset shows the variation of the layer spacing $d_1$ with temperature, obtained from the XRD data.}
\label{figure_XRD}
\end{figure}

The X-ray diffraction (XRD) studies were conducted at different temperatures to investigate the
molecular organization in the observed mesophase. The X-ray intensity profile as a function of scattering wave vector in the liquid crystalline phase of the sample is shown in Fig.~{\ref{figure_XRD}}. Three 
sharp peaks in the small-angle region centred at about 0.114 $\mathrm{\AA^{-1}}$, 
0.228 $\mathrm{\AA^{-1}}$ and 0.341 $\mathrm{\AA^{-1}}$ respectively were observed. These 
sharp peaks are in the ratio of 1:2:3 of the wave vectors. These peaks clearly indicate a 
smectic-like layered structure of the mesophase with layer spacing $d_1 = 55.1 \mathrm{\AA}$. 
In addition, a single diffused peak centred at about 1.360 $\mathrm{\AA^{-1}}$  was observed 
in the wide-angle region.  The diffused peak indicates an in-plane liquid-like ordering of the 
molecules with an average intermolecular separation of 4.7 $\mathrm{\AA}$. The x-ray intensity profiles 
 at different temperatures of the smectic phase are found to be identical and superimpose on each other. Therefore, the layer spacing of the smectic phase remains constant with temperature as shown in the inset 
of Fig.\ref{figure_XRD}. The measured layer spacing $d_1=55.1 \mathrm{\AA}$ is almost equal to the molecular length $l=53.8 \mathrm{\AA}$. Therefore, the observed mesophase is of Smectic-A type without much intercalation between the molecules in adjacent layers.
\label{subsection_XRD}
\subsection{\textbf{Current response of the planar aligned sample}}
\begin{figure}[!t]
\centering
\includegraphics[width=0.75\linewidth]{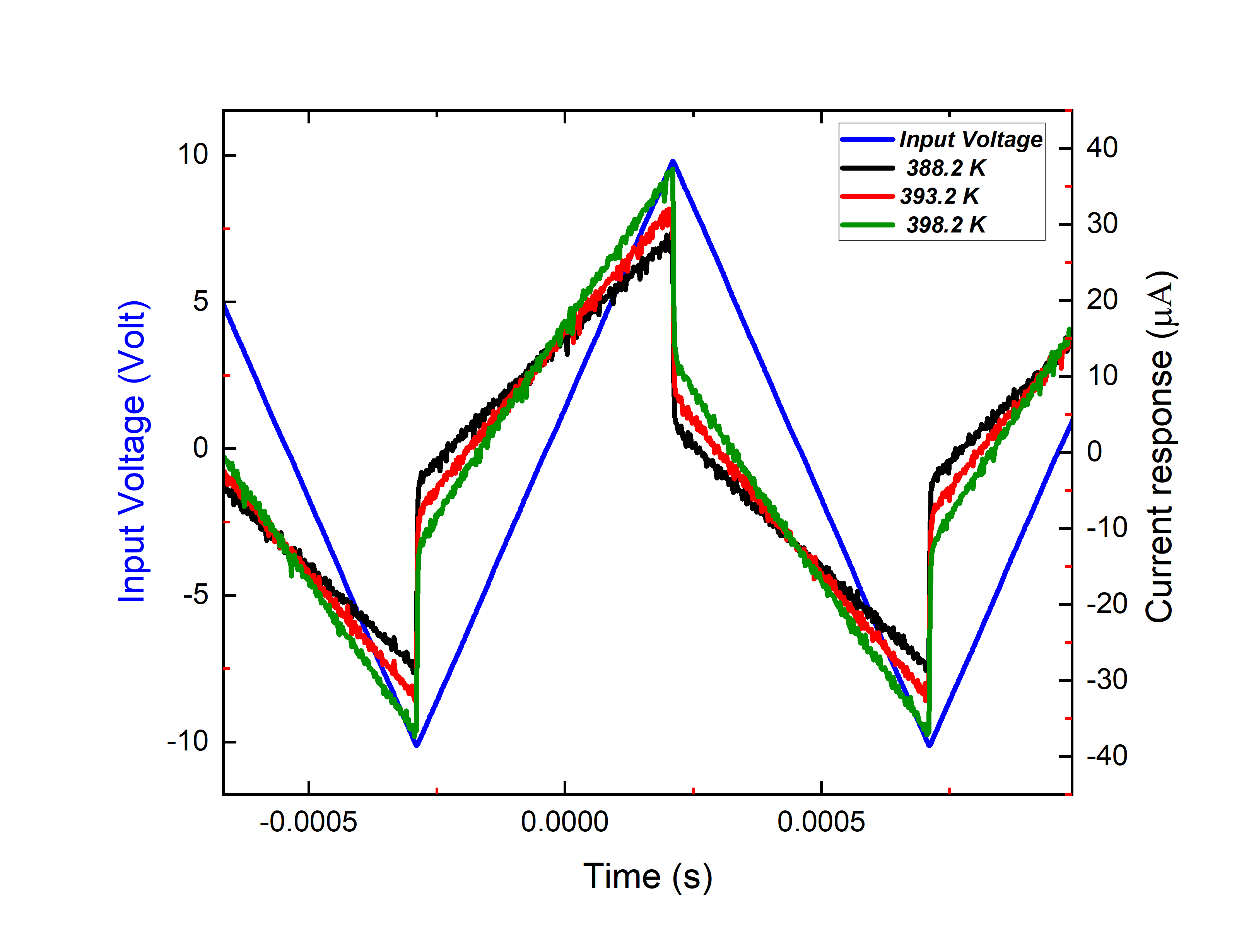}
\caption{The triangular wave voltage current response of the planar aligned sample at three different temperatures in the SmA$_b$ phase. The 
absence of polarisation reversal current peaks indicates the apolar nature of the SmA$_b$ phase.}
\label{figure_Cuurent_response}
\end{figure}

The current response using the triangular wave voltage technique was measured to investigate 
the electric polarisation in the planar aligned smectic-A phase of the sample.
Fig.~\ref{figure_Cuurent_response} shows the current response of the sample at three different
temperatures in the Smectic-A phase. The absence of current peaks corresponding to polarisation
reversal indicates that there is no spontaneous polarisation in the smectic layers. Hence, the
smectic-A phase is apolar in nature. 
\subsection{POM Studies}
As mentioned earlier, the sample in the smectic-A phase was found to align homeotropically 
in a custom made cell made of two ITO-coated glass plates without any treatment for molecular
alignment.
\begin{figure}[!t]
\centering
\includegraphics[width=1\linewidth]{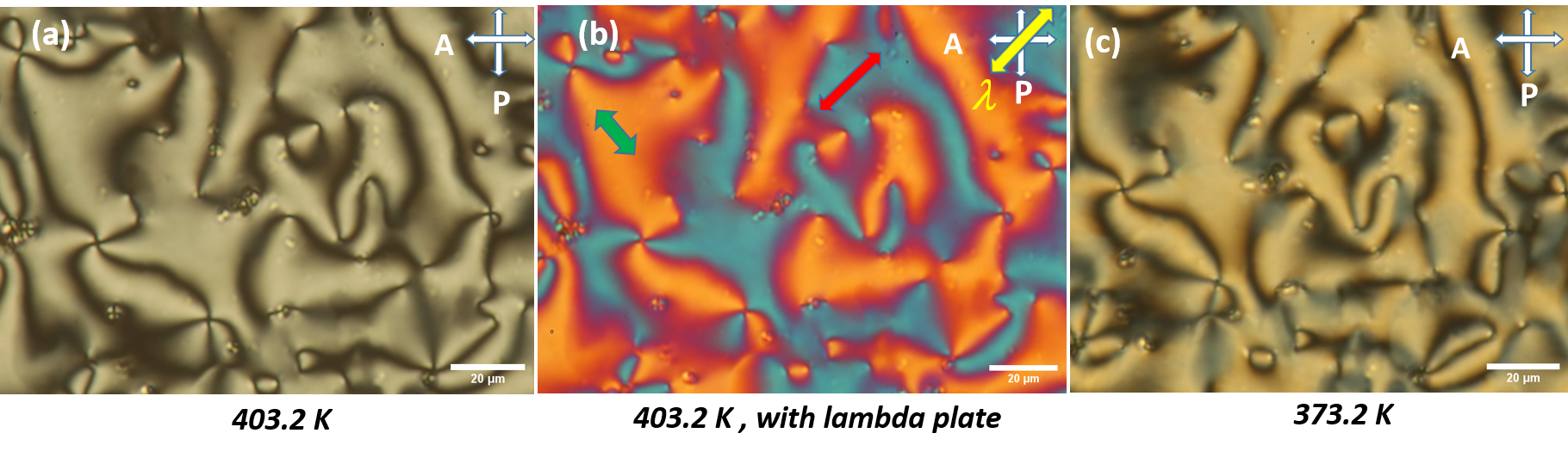}
\caption{POM textures of the homeotropically aligned region of an 8.0 $\mu$m thick sample in a custom made LC cell. (a) The Schlieren texture of the SmA$_b$ phase at 403.2 K. (b) The texture shown in (a) when a $\lambda$ plate of wavelength 530 nm is inserted in the optical path.
(c) The texture at 373.2 K of the SmA$_b$ phase. The white arrows denote the orientation of the cross-polarisers, and the yellow arrow at 45 degrees to the polarisers denotes the slow axis
of the lambda plate. The scale bar represents a length of 20 $\mu$m.}
\label{figure_pom_custom}
\end{figure}
The POM observations of a homeotropically aligned sample of thickness about 8.0 $\mu$m were 
performed during cooling from the isotropic phase. At about 406.2 K, the sample shows a 
sharp transition from isotropic to Smectic-A phase. The sample in Smectic-A phase aligns 
homeotropically in most of the region. Some portion of the sample also shows fan-like
textures with planar alignment. In the homeotropically aligned region of the smectic-A phase, 
a schlieren texture with numerous defects having both two and four brushes
were observed as shown in Fig.~\ref{figure_pom_custom}(a). The two and four brush defects 
correspond to strength $\pm 1/2$ and $\pm 1$ respectively.
The presence of $\pm 1/2$ defects in the homeotropically aligned Smectic-A phase clearly indicates
the biaxial nature of this phase\cite{pratibha2000science, smalyukh2005selective}. The $\pm 1/2$ defects are not expected in the transversely polar ferroelectric Smectic-A ($SmAP_F$) phase of bent-core molecules. Further, the polarization reversal current measurements described above indicate the absence of layer polarization in the observed Smectic-A phase. Therefore, the XRD, POM and polarization reversal current measurements clearly show that the Smectic-A phase is biaxial in nature, i.e., the SmA$_b$ phase.

Fig.~\ref{figure_pom_custom}(b) shows the Schlieren texture with the insertion of a lambda plate (530 nm) in the optical path of the microscope. The blue (orange) region of the texture shows that the secondary director of the SmA$_b$ phase is parallel (perpendicular) to the slow axis of the lambda plate. The defect configuration in the schlerien texture did not change on decreasing the temperature, but the colour changed from greyish to yellowish as shown in Fig.~\ref{figure_pom_custom}(c). This change in colour indicates that the effective birefringence in homeotropic alignment of the 
SmA$_b$ phase increases with decreasing temperature. 

\begin{figure}[ht] 
\centering
\includegraphics[width=1\linewidth]{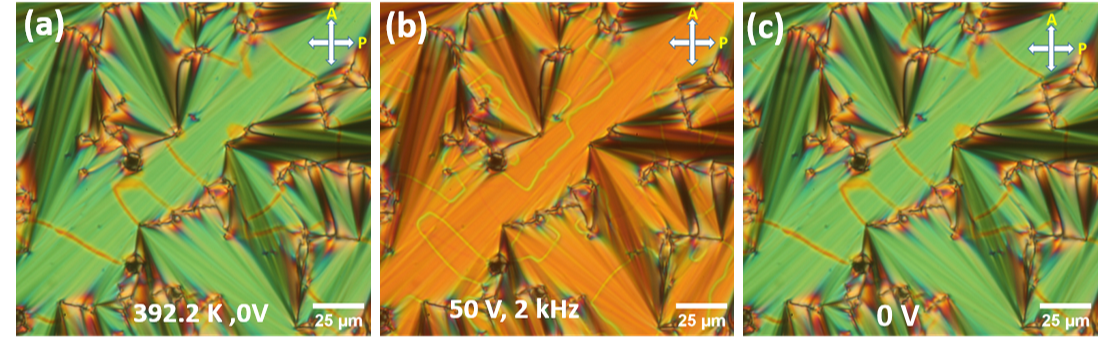}
\caption{ POM textures of the planner-aligned SmA$_b$ phase at 392.2 K of thickness 9$\mu$m in a commercial LC cell. The fan textures between cross-polariser (denoted by white cross arrows), (a) in the absence of applied voltage, 
(b) in the presence of an applied sinusoidal voltage of amplitude 50 volt and frequency 2 kHz. (c) after switching off the applied voltage. The scale bar represents a length of 20$\mu$m.}
\label{fig_9mp_pom}
\end{figure}

The POM observations for planar aligned samples were carried out in 5 $\mu$m commercial 
LC cells.
\begin{figure}[t] 
\centering
\includegraphics[width=1\linewidth]{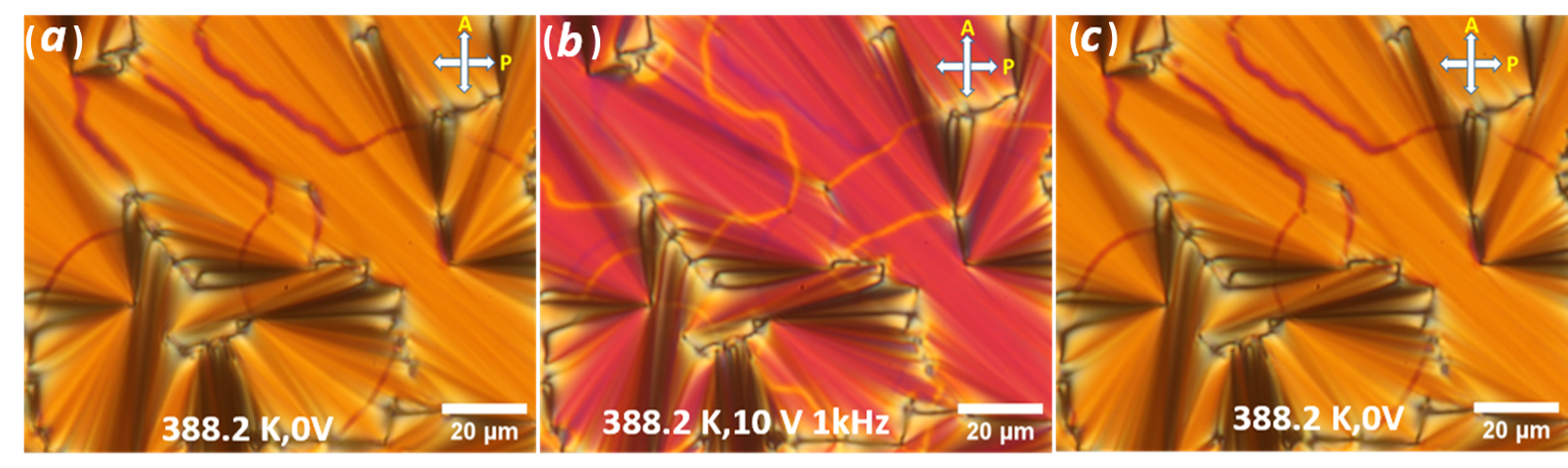}
\caption{POM textures of the planner-aligned SmA$_b$ phase at 388.2 K of thickness 5$\mu$m in a commercial LC cell.
The focal conic fan texture between crossed polarisers (a) in the absence of applied voltage, 
(b) in the presence of an applied sinusoidal voltage of amplitude 10 volt and frequency 1 kHz. (c) after switching off the applied voltage. The scale bar represents a length of 20$\mu$m. 
}
\label{figure_pom_5mp}
\end{figure}
The focal conic fan-like texture (Fig.~\ref{figure_pom_5mp}(a)) typical of the smectic phase 
was observed at 388.2~K in the absence of an applied electric field. The fan regions in which the smectic layer normal is oriented either parallel or perpendicular to the polarizer appear dark. This 
observation indicates that the major principal axis of the optical indicatrix is 
parallel to the layer normal, confirming the smectic-A type phase. Interestingly, it is found that the colour 
of the texture changes reversibly with the application of an electric field above a threshold voltage 
(see Fig.~\ref{figure_pom_5mp}(b)). The zero-field texture can be recovered by switching off the electric field as shown in Fig.~\ref{figure_pom_5mp}(c). The field-induced colour change, while preserving the overall texture, can be attributed to a variation in birefringence. Thus, this transition can be interpreted as a Fréedericksz-like reorientation of the secondary director in the biaxial SmA$_b$ phase~\cite{meyer2021freedericksz,walker2021remarkable}. Similar results were also observed using a 9 $\mu$m thick planar aligned sample (See Fig.~\ref{fig_9mp_pom}). These observations clearly confirm the biaxiality of the SmA$_b$ phase. 
\label{subsection_POM}

We further investigated the temperature dependence of the effective birefringence of a 5 $\mu$m thick planar-aligned sample in the SmA$_b$ phase. The effective birefringence was evaluated by measuring the transmitted intensity through the sample between crossed polarizers using polarizing optical microscope. For a homogeneous planar aligned liquid crystalline sample kept between crossed polarizers, the intensity of transmitted light can be written as
\begin{equation}\label{Int_for_biref_main_eq}
\textbf{$I =\frac{I_{0}}{2} \sin ^2(2\psi )(1-\cos\Delta\Phi) + I_{d},$}
\centering
\end{equation}
where $I_{0}$ is the intensity of the incident light, $\psi$ is the angle between the polariser and the local primary director $\textbf{\^n}$ and $I_{d}$ is the intensity under dark conditions measured in the isotropic phase of the sample between crossed polarisers. The phase difference  
\begin{equation}\label{phase_for_biref_main_eq}
\textbf{$\Delta\Phi$ = $\frac{2\pi \Delta n d}{\lambda}$}
\centering
\end{equation}
where d is the sample thickness, $\lambda$ is the wavelength of the incident light and $\Delta n$
is the effective birefringence of the sample. During the experiment, the sample was placed in such a way that the region of interest had maximum brightness, i.e. $\psi= \pi/4 $. Consequently Eq.~\ref{Int_for_biref_main_eq} becomes 
\begin{equation}
\textbf{$I =\frac{I_{0}}{2}(1-\cos\Delta\Phi) + I_{d}$}
\centering
\label{Int_for_biref_reduced_eq}
\end{equation}
\begin{figure}[!ht]      
\centering
\includegraphics[width= 0.75\linewidth]{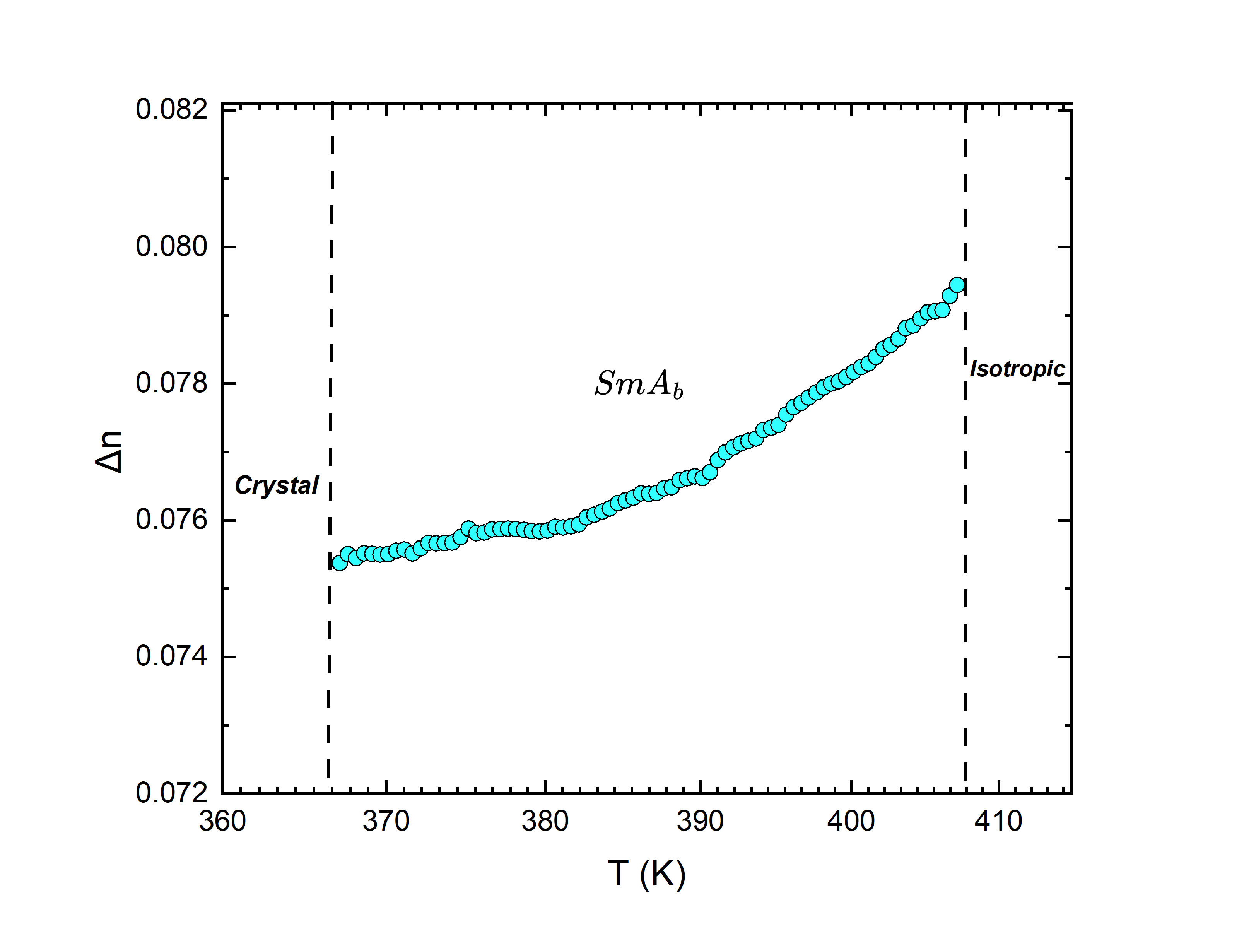}
\caption{The variation of the effective birefringence $\Delta n$ of the planar-aligned SmA$_b$ phase with  temperature.}
\label{fig:Biref_vs_T}
\end{figure}
Thus from Eq.~\ref{Int_for_biref_reduced_eq}, the minimum and maximum intensities are given by$I_{min} = I_{d}$ and $I_{max} =I_{0}+ I_{d}$ respectively. In our experiments, $I_{min}$ and $I_{max}$ were measured while the sample was in the isotropic phase with the polariser and analyzer perpendicular and parallel, respectively. Using these values, the effective birefringence $\Delta n$ of the planar aligned SmA$_b$ phase was calculated from Eq.~\ref{Int_for_biref_reduced_eq} as a function of temperature. 

The optical indicatrix of the SmA$_b$ phase is characterized by three mutually perpendicular principal axes associated with three distinct principal refractive indices, $n_1 ,n_2$  and $n_3$. The refractive index $n_3$ is assumed to be along the primary director  $\textbf{\^n}$, which is parallel to the layer normal of the SmA$_b$ phase. While  $n_2$ is along the secondary director \textbf{\^m}, which is parallel to the average orientation direction of the planes of the bent-core molecules in the SmA$_b$ layers. The least refractive index $n_1$ is along the axis $\textbf{\^l}$, which is perpendicular to both the director $\textbf{\^n}$ and $\textbf{\^m}$. It is reasonable to assume that $n_3 >n_2>n_1$ for the observed SmA$_b$ phase. For the planar aligned sample, the measured effective birefringence $\Delta n = n_3 - n_2$ as a function of temperature in the SmA$_b$ phase is shown in Fig.~\ref{fig:Biref_vs_T}. The effective birefringence $\Delta n$ exhibits an anomalous increase with increasing temperature, in contrast to the behaviour of uniaxial SmA phase. Such a trend can be naturally explained by the biaxial ordering present in the SmA$_b$ phase. The textures of the homeotropically aligned sample (Fig.~\ref{figure_pom_custom}(a), (c)) indicate that the effective optical biaxiality $\delta n = n_2 - n_1$ in the SmA$_b$ phase decreases with increasing temperature. Thus, this decrease in optical biaxiality $\delta n$  gives rise to decrease in $n_2$ and increase in $n_1$ with increasing temperature. Assuming that the refractive index $n_3$ does not vary appreciably with temperature, the decrease of $n_2$ gives rise to the increase of the effective birefringence $\Delta n$  with increasing temperature as shown in Fig.~\ref{fig:Biref_vs_T}.

\label{subsection_BIREFRINGENCE_T}
\subsection{\textbf{Dielectric studies}}
\begin{figure}[ht] 
\centering
\includegraphics[width=1\linewidth]{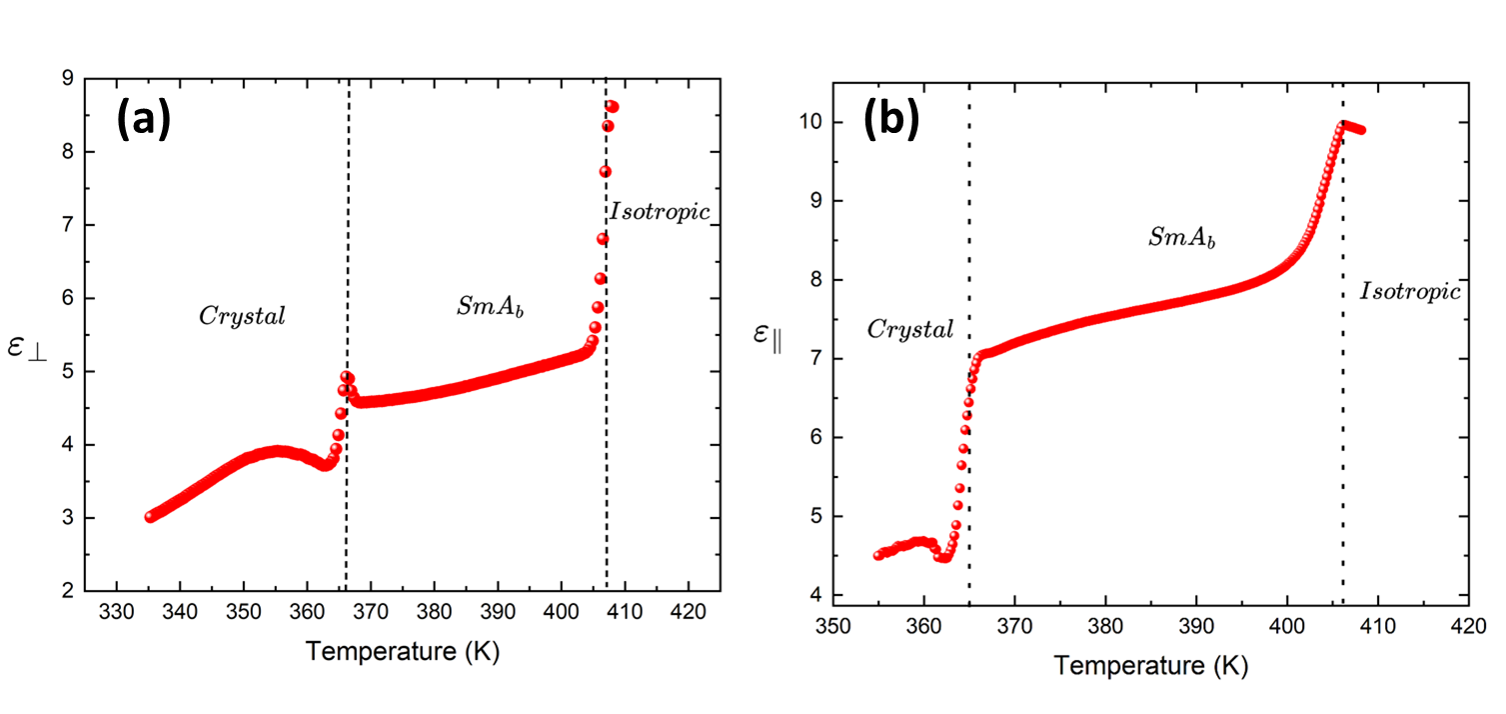}
\caption{The variation of effective dielectric constant of the SmA$_b$ phase, (a) $\varepsilon_{\perp}$ as a function of temperature, measured on a planar-aligned 5 $\mu$m thick sample, (b) $\varepsilon_{\parallel}$ as a function of temperature of a homeotropic sample of thickness of 8 $\mu$m in a custom made LC cell. The sample exhibits positive dielectric anisotropy $\Delta \varepsilon \approx 2$.}
\label{figure_dielectric_vs_temp}
\end{figure}

The dielectric properties of the sample CBBO20 as a function of temperature and 
frequency of the probing electric field were measured for both planar and homeotropic 
alignments. The temperature variation of effective $\varepsilon_{\perp}$ and $\varepsilon_{\parallel}$ of the sample 
for planar and homeotropic alignment are shown in Fig.~\ref{figure_dielectric_vs_temp}(a), (b) respectively. The variation of the dielectric constants with temperature clearly detects the transitions between the observed phases, which agree well with the transition temperatures found in the POM studies. 
The effective dielectric permittivity $\varepsilon_{\perp}$  decreases sharply at the isotropic–SmA$_b$ phase transition. Upon further cooling within the SmA$_b$ phase, it exhibits a gradual decrease with temperature. At the SmA$_b$–crystal transition, $\varepsilon_{\perp}$ displays a small peak followed by a dip before entering the crystalline phase.
The measured low value of the dielectric constant in the planar aligned sample indicates the absence of layer polarization in the SmA$_b$ phase confirming the apolar nature of this phase. 

\begin{figure}[ht] 
\centering
\includegraphics[width=1\linewidth]{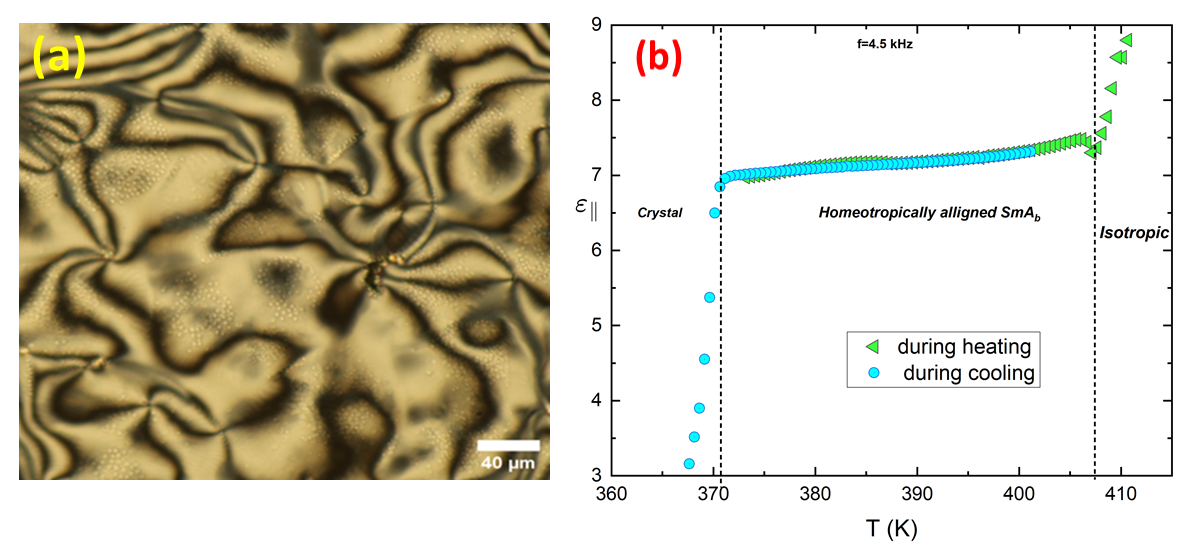}
\caption{(a) The schlieren texture of homeotropically aligned SmA$_b$ phase obtained on cooling a sample of thickness 9 $\mu m$ from its isotropic phase with  the application of a  sinusoidal voltage of r.m.s amplitude 30 V and frequency 20 kHz. (b) The variation of dielectric constant with temperature of the same sample during heating and cooling.}
\label{fig:9mp_homeo_diel_vs_T}
\end{figure}

For the homeotropically aligned sample, the effective $\varepsilon_{\parallel}$ decreases sharply across the isotropic to the SmA$_b$ phase, then decreases gradually in the SmA$_b$ phase and drops sharply in the crystal phase with decreasing temperature as shown in Fig.~\ref{figure_dielectric_vs_temp} (b). The measured $\varepsilon_{\parallel}$ is larger than $\varepsilon_{\perp}$ in the SmA$_b$ phase indicating  positive dielectric anisotropy $\Delta\varepsilon \approx 2$. The positive dielectric anisotropy is expected as the molecule possesses a strong dipole moment along its long axis as found in the energy-minimised structure (see Fig.~\ref{figure_DSC}(a)).
Similar measurements of $\varepsilon_{\parallel}$ with temperature were also carried out in 
a 9 $\mu$m planar cell in which the sample was aligned homeotropically as shown in Fig.~\ref{fig:9mp_homeo_diel_vs_T}(a) using a high electric field as discussed in Sec.~\ref{Section__Expt}. The results shown in Fig.~\ref{fig:9mp_homeo_diel_vs_T}(b) agree well with that in Fig.~\ref{figure_dielectric_vs_temp}(b).

\begin{figure}[ht] 
\centering
\includegraphics[width= 1\linewidth]{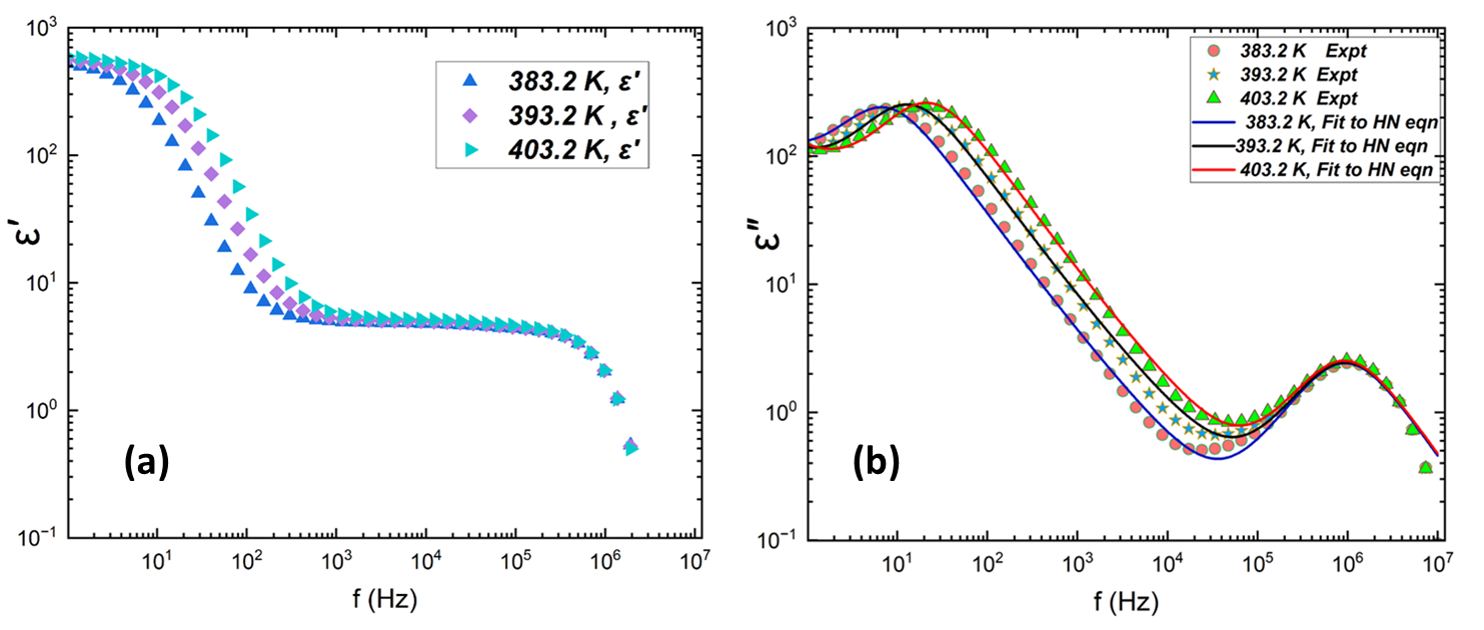}
\caption{The variation of (a) real part $\varepsilon'$ and (b) imaginary part $\varepsilon''$ of the complex dielectric function with frequency of a planar aligned 9 $\mu m$ thick sample in the SmA$_b$ phase at different temperatures. The solid lines in (b) denote the best fit to the experimental data using the Havriliak-Negami equation.}
\label{figure_planar_diel}
\end{figure}

The dielectric relaxation behaviour of both planar and homeotropically aligned samples was
investigated by measuring the complex dielectric constant as a function of frequency of the 
probing electric field in the range of 1 Hz to 10 MHz at different temperatures. The 
frequency-dependent complex dielectric function of the sample can be written
as~\cite{kremer2002broadband}
\begin{equation}
\label{dielectric_fuction_equation}
\varepsilon^{*}(f)=\varepsilon'(f)-j\varepsilon''(f)
\end{equation}
where $f$ is the frequency of the applied field, $\varepsilon'$ and $\varepsilon''$ are real 
and imaginary parts of the complex dielectric function, respectively. 
The frequency dependence of $\varepsilon'$ and $\varepsilon''$ at different temperatures 
for the planer aligned Sm$A_{b}$ phase are shown in Fig.~\ref{figure_planar_diel}(a), (b) respectively. Two relaxation processes corresponding to the peaks in the imaginary part of the dielectric function were observed for the planar aligned sample. 
\begin{figure}[ht] 
\centering
\includegraphics[width=1\linewidth]{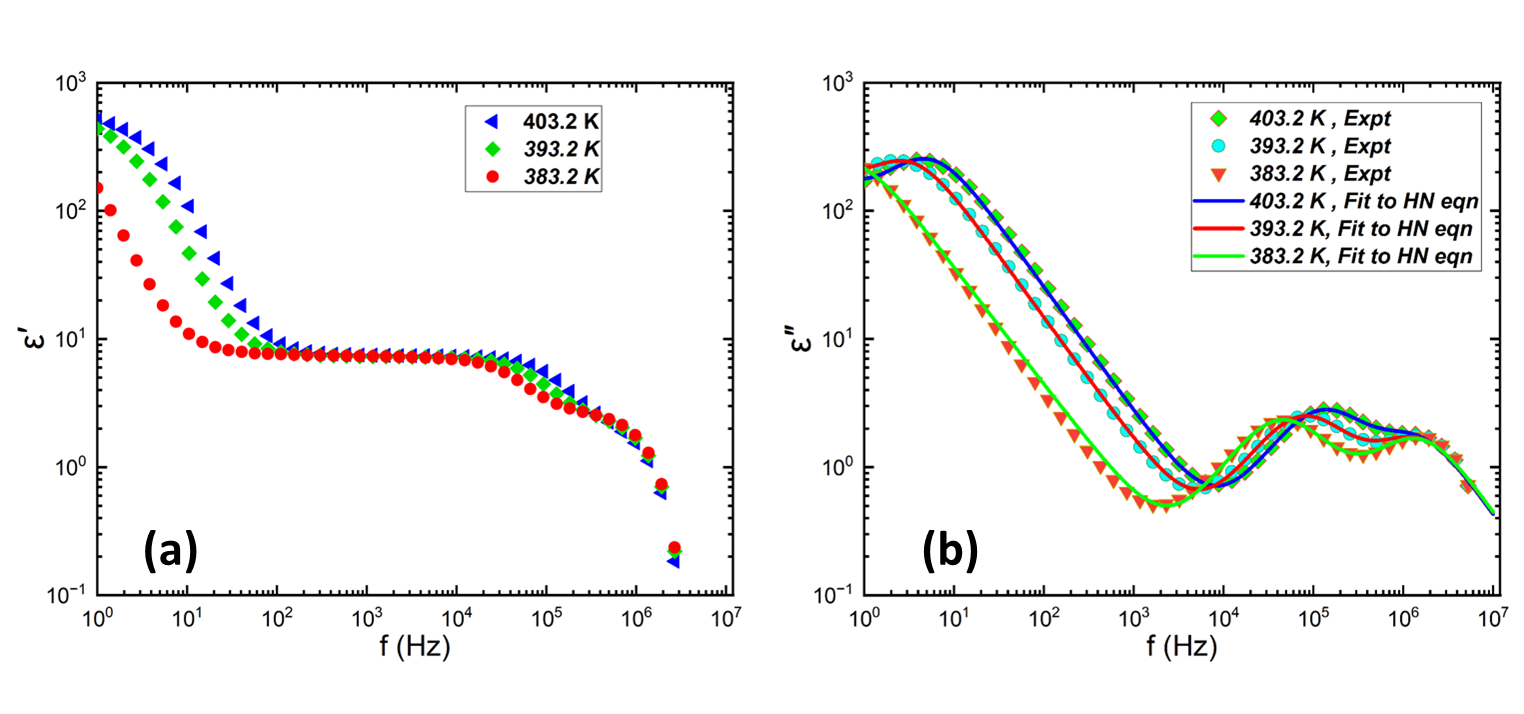}        
\caption{The variation of (a) real part $\varepsilon'$ and (b) imaginary part $\varepsilon''$ of the complex dielectric function with frequency of a homeotropically aligned 9 $\mu m$ thick sample in the SmA$_b$ phase at different temperatures. The solid lines in (b) denote the best fit to the experimental data using the Havriliak-Negami equation.}
\label{figure_homeotropic_diel}
\end{figure}
Fig.~\ref{figure_homeotropic_diel}(a), (b) shows the frequency dependence of $\varepsilon'$ and
$\varepsilon''$ at different temperatures for the homeotropically aligned SmA$_b$ phase, respectively. 
Three relaxation peaks of $\varepsilon''$ as a function of frequency were observed 
for the homeotropically aligned SmA$_b$ phase.

\begin{figure}[ht] 
\centering
\includegraphics[width=1\linewidth]{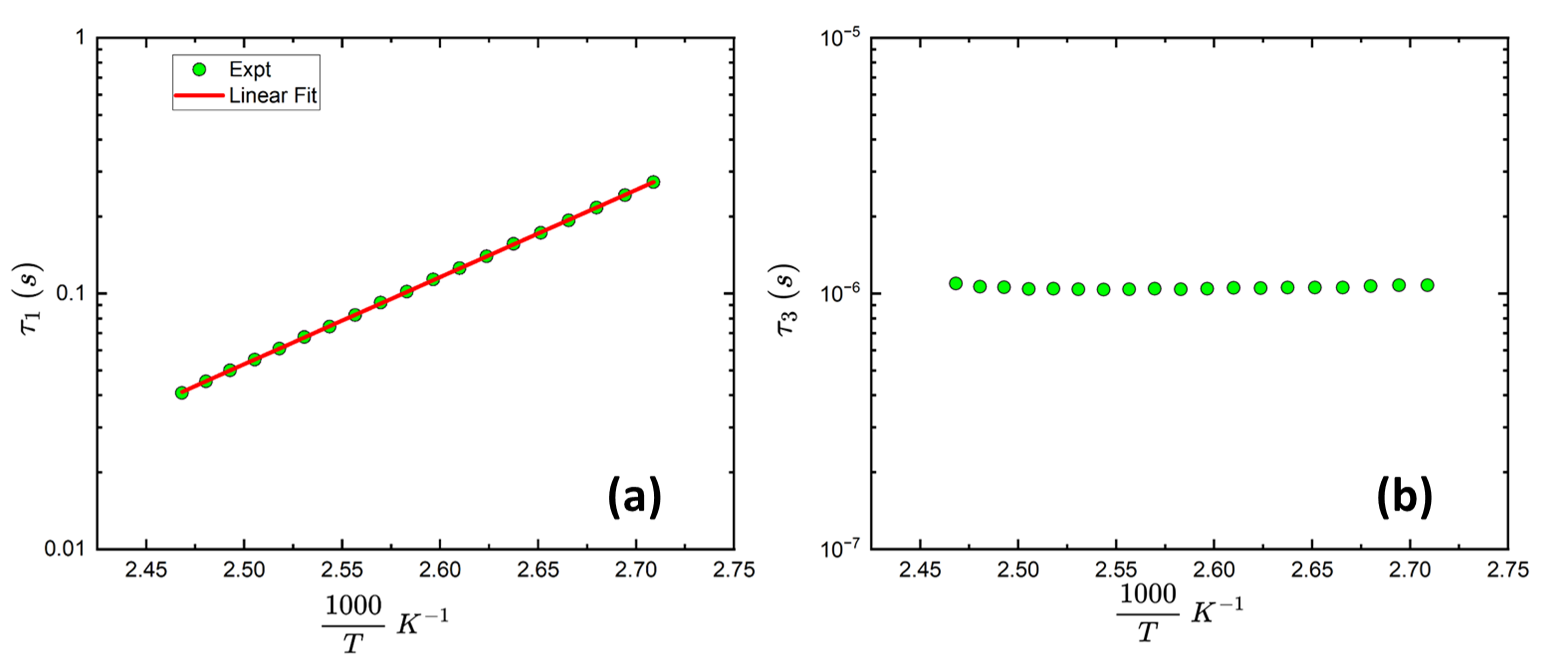}
\caption{ The Variation of the  relaxation time  with inverse temperature of a 9 $\mu m$ thick planar aligned sample for (a) the low-frequency ionic mode  (b ) the high-frequency mode  due to the ITO coating .
The solid line in (a) denotes the  best fit to the experimental data using the Arrhenius equation.}
\label{figure_plnr_Log(Tau)_vs_invT}
\end{figure}

To analyze the dielectric spectra, the imaginary part of the complex dielectric functions were fitted using the Havriliak-Negami (HN) equation\cite{havriliak1966complex,havriliak1967complex}. This empirical equation expresses the angular frequency dependent complex dielectric function $\varepsilon^*(\omega)$ in terms of the various relaxation modes as given by
\begin{equation}
\label{HN_equation}
\varepsilon^{*}(\omega)-\varepsilon_{\infty}
=
\sum_{j=1}^{n}
\frac{\Delta\varepsilon_{j}}
{\left[1+\left(i\omega\tau_{j}\right)^{\alpha_{j}}\right]^{\beta_{j}}}
-\frac{i\,\sigma_{\mathrm{dc}}}
{\varepsilon_{0}\omega^{s}}
\end{equation}
where $\Delta\varepsilon_{j}$ represents the dielectric strength of the j-th relaxation mode,
$\tau_{j}$ represents the corresponding relaxation time. The constants $\alpha_{j}$ 
($0<\alpha_{j}\le 1$) and $\beta_{j}$ ($0<\alpha_{j}\beta_{j}\le 1$) are the shape parameters 
and describe broadness and asymmetry of the dielectric loss spectra, respectively.  
The last term on the RHS of Eq.~\ref{HN_equation} arises due to the dc conductivity
$\sigma_{\mathrm{dc}}$ of the sample. The constant $s$ ($0< s \le 1$) is a fitting parameter and its value is $1$ for purely ohmic conduction. The influence of electrode polarization can give rise to $s<1$ \cite{kremer2002broadband}.
The constant $\varepsilon_{\infty}$ is the high-frequency limit of the dielectric constant and $\varepsilon_{0}$ is the permittivity of free space. The HN equation reduces to the Cole-Cole function when $\beta$=1 and to the Cole-Davidson equation when $\alpha$=1. The relaxation process is classified as Debye type when both $\alpha$=1 and $\beta$=1. 
The dielectric spectra corresponding to both planar and homeotropic alignments of the 
sample were successfully fitted using  Eq.~\ref{HN_equation}. The fitting of
$\varepsilon''$ for both planar and homeotropically aligned sample are shown by the solid lines in Fig.~\ref{figure_planar_diel}(b) and 
Fig.~\ref{figure_homeotropic_diel}(b), respectively.
All the observed dielectric relaxation processes are found to be of Debye type in the SmA$_b$ phase.
The relaxation peak in the low-frequency region has 
greater dielectric strength and it shifts towards the higher frequencies with increasing temperature.
This low-frequency relaxation mode is  attributed to  small concentration of ionic charges present in the sample. The dielectric strength and relaxation frequency of the mode
corresponding to the peak at about $1$ MHz remain constant 
throughout the temperature range of study for both planar and homeotropically 
aligned samples. This mode is present even in the isotropic phase and can
be attributed  to the finite sheet resistance of the ITO coating on the glass plates of the LC cells \cite{ARoy2024PRE}. The variation of the relaxation times with
inverse temperature of the low frequency mode due to the ionic charge impurities is found to be of Arrhenius type as shown in  Fig.~\ref{figure_plnr_Log(Tau)_vs_invT}(a). The relaxation time of the high frequency ITO mode remains constant with temperature (See Fig.~\ref{figure_plnr_Log(Tau)_vs_invT}(b)). Both these dielectric modes are not associated with the SmA$_b$ phase and will not be discussed further. 

\begin{figure}[ht] 
\centering
\includegraphics[width= 1\linewidth]{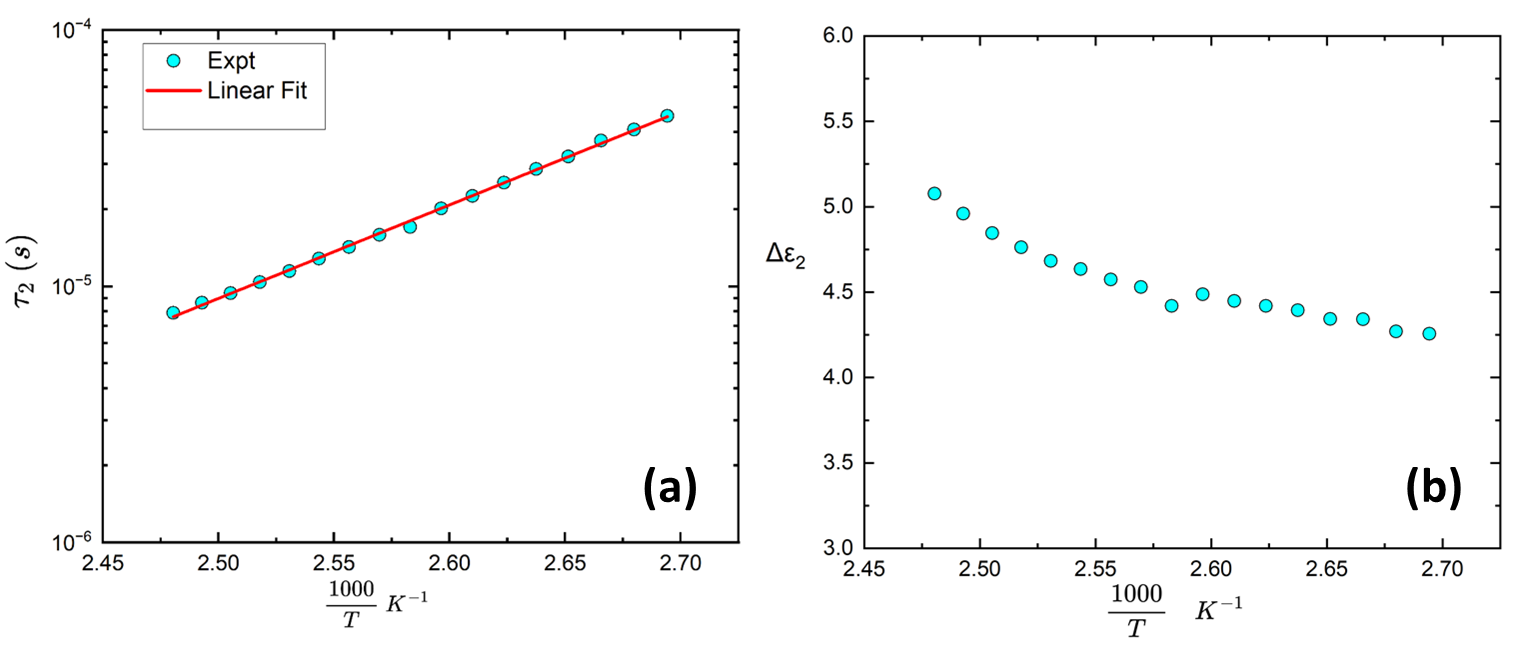}
\caption{ The variation of (a )relaxation time  and (b) dielectric strength, as a function of inverse temperature of the dielectric relaxation mode associated with the flipping of the molecules in the smectic layers observed in a 9 $\mu$m thick homeotropically aligned SmA$_b$ phase. The solid line is the best fit to the experimental data using the Arrhenius equation.}
\label{figure:homeo_2ndpeak_Log(Tau)_and_Eps2_vs_invT}
\end{figure}

For homeotropic alignment of the sample, three dielectric relaxation modes were observed.
In addition to the modes associated with the ionic impurities and ITO coating, another mode with intermediate relaxation time is found in the SmA$_b$ phase.
Fig.~\ref{figure:homeo_2ndpeak_Log(Tau)_and_Eps2_vs_invT}(a), (b) shows the variation of the relaxation 
time and dielectric strength of this intermediate dielectric relaxation mode with 
inverse temperature, respectively. The linear variation of the
logarithm of the relaxation time with inverse temperature indicates that the mode is of 
Arrhenius type with activation energy of 0.31 eV. This mode can be associated with the flipping of the
strongly polar molecules about their short axis in the smectic layers. The increase in the viscosity of the medium with decreasing temperature hinders such molecular flipping, resulting in an increase in the relaxation time and decrease in dielectric strength, consistent with the experimental observations. This mode was not observed in the planar aligned sample as the applied field is in the plane of the smectic layers which
can not excite such molecular flipping . In the planar sample, the mode corresponding to the rotation about the long axis of the molecules is expected.  However, the relaxation frequency of this mode is generally high and perhaps it lies beyond the frequency range of our experimental setup.
\label{subsection_DIELECTRIC}

\begin{figure}[ht]
\centering
\includegraphics[width=0.75\linewidth]{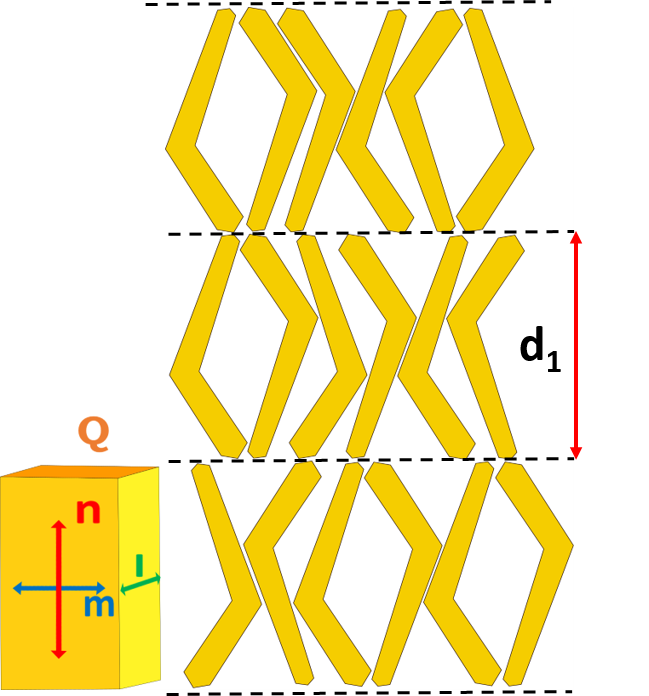}
\caption{The schematic representation of the proposed structural model of the observed Sm$A_b$ phase of bent core molecules. The long axes of the molecules align on average along the layer normal with the  layer  spacing $d_1 $  close to the molecular length.  The biaxial orientational ordering  tensor \textbf{Q} illustrates the mutually orthogonal directors  $\textbf{\^l}$, $\textbf{\^m}$ and $\textbf{\^n}$, where  $\textbf{\^n}$ is parallel to the layer normal,  $\textbf{\^m}$ lies in the layer plane, and $\textbf{\^l = \^n × \^m}$.}
\label{figure_layer_spacing_model}  
\end{figure}

Based on our experimental results, the proposed molecular organization in the observed SmA$_b$ phase is schematically shown in Fig.~\ref{figure_layer_spacing_model}. The layer spacing $d_1$ is close to the molecular length of the bent core molecules with the primary director $\textbf{\^n}$ parallel to the layer normal. The arrow directions of the bent core molecules align themselves along the secondary director $\textbf{\^m}$ and are statistically equivalent along  $\textbf{\^m}$ and $-\textbf{\^m}$, leading to the absence of net layer polarization.
The ordering tensor $\textbf{Q}$ with three mutually perpendicular principal axes $\textbf{\^l}$, $\textbf{\^m}$ and $\textbf{\^n}$, is also illustrated in Fig.~\ref{figure_layer_spacing_model}. It should be noted that the biaxial ordering described by the unit vectors  $\textbf{\^l}$,  $\textbf{\^m}$ and  $\textbf{\^n}$ is invariant under the inversion of  coordinate system. The bent core molecules usually pack efficiently giving rise to polar order in the layer. The relatively large bend angle of the asymmetric molecules in the studied compound may favor such biaxial ordering within the smectic layers. 

\section{Conclusion}
We report experimental studies on a pure compound consisting of bent-core banana-shaped molecules, 
which exhibits an enantiotropic biaxial Smectic-A phase with no net layer polarization. The XRD studies show that the layer spacing is close to the molecular length and remains constant throughout the temperature range of the SmA$_b$ phase. The POM studies further confirmed the biaxiality of the  SmA$_b$ phase. The polarization reversal current measurements indicate the
absence of spontaneous polarization in the biaxial smectic layers. A dielectric relaxation 
mode associated with the reorientation of molecules about their short molecular axis was 
observed in the SmA$_b$ phase. This mode is detected only in the homeotropic alignment of 
the sample. The relaxation frequency decreases rapidly with decreasing temperature and 
follows the Arrhenius relation. Interestingly, the SmA$_b$ phase is found to exhibit an electro-optic effect which resembles the Fréedericksz transition in nematic phase. The details of the electro-optic effect in the SmA$_b$ phase will be publish elsewhere.  
\section{ACKNOWLEDGMENTS}
We gratefully acknowledge Ms Vasudha K. N. for her assistance in acquiring the DSC and XRD data. 
\nocite{*}
\bibliography{reference}

\end{document}